\documentclass [12pt]{article}
\usepackage[a4paper, margin=1in]{geometry}

\usepackage{graphicx}		
\usepackage{amsmath}				
\usepackage{bm}						
\usepackage{float}
\usepackage{natbib}					
\usepackage{amssymb}				
\usepackage{longtable}				
\usepackage{verbatim}				
\usepackage{kbordermatrix}
\usepackage{url}
\usepackage{lscape}
\usepackage{xcolor}

\usepackage[margin=20pt, font=small, labelfont=bf]{caption}

\title{}
\author{}

\begin{document}

\title{Optimal Consumption, Investment and Housing with Means-tested Public Pension in Retirement}
\author{Johan G. Andr\'{e}asson \footnote{School of Mathematical and Physical Sciences, University of Technology, Sydney, Broadway, PO Box 123, NSW 2007, Australia; email: andreasson.johan@uts.edu.au} , Pavel V. Shevchenko \footnote{CSIRO, Australia; University of Technology, Sydney, Australia; email: pavel.shevchenko@csiro.au} , Alex Novikov \footnote{School of Mathematical and Physical Sciences, University of Technology, Sydney, Australia; email: alex.novikov@uts.edu.au}}
\date{\today}
\maketitle

\begin{abstract}
In this paper, we develop an expected utility model for the retirement behavior in the decumulation phase of Australian retirees with sequential family status subject to consumption, housing, investment, bequest and government provided means-tested Age Pension. We account for mortality risk and risky investment assets, and introduce a health proxy to capture the decreasing level of consumption for older retirees. Then we find optimal housing at retirement, and optimal consumption and optimal risky asset allocation depending on age and wealth. The model is solved numerically as a stochastic control problem, and is calibrated using the maximum likelihood method on empirical data of consumption and housing from the Australian Bureau of Statistics 2009-2010 Survey. The model fits the characteristics of the data well to explain the behavior of Australian retirees. The key findings are the following: first, the optimal policy is highly sensitive to means-tested Age Pension early in retirement but this sensitivity fades with age. Secondly, the allocation to risky assets shows a complex relationship with the means-tested Age Pension that disappears once minimum withdrawal rules are enforced. As a general rule, when wealth decreases the proportion allocated to risky assets increases, due to the Age Pension working as a buffer against investment losses. Finally, couples can be more aggressive with risky allocations due to their longer life expectancy compared with singles.  \\


\end{abstract}

\newpage

\section{Introduction}
The global shift from a defined-benefit to a defined-contribution pension system transfers risk from the corporate sector to households, primarily via the investment and withdrawal decision of pension assets. Australia's move has already accumulated superannuation\footnote{In Australia, the arrangement for people to accumulate funds for income in retirement is referred to as \emph{superannuation}.} assets of \$2.02 trillion dollars \citep{ASFA2015}, making Australia the 4th largest pension fund pool in the world. For retirees, the main difference is that instead of being provided a monthly benefit as they were before, they now receive a lump sum at retirement and are responsible for managing this wealth throughout their lives. The retirees face multiple risks including mortality, longevity and investment risks, as well as regulatory risk such as changes in policies and government provided Age Pension entitlements, which are harder to account for. The long term effects of this new pension system are not yet known, and there is limited knowledge amongst retirees and advisors regarding how to consume and manage the assets. This results in confusion for many retirees \citep{Agnew2013}.

Once retired, Australian retirees tend to keep the same proportion of risky assets as before retirement, even though the exposure decreases with age \citep{SpicerEtAl2013}. With no labor income these assets are subject to sequencing risk\footnote{Sequencing risk refers to unfortunate timing of cash flow in the investment portfolio. Drawdown after negative investment returns will have a greater impact on the long term growth of the portfolio compared with drawdown after positive investment returns.} which is difficult to avoid in the decumulation phase, especially early in retirement \citep{Kingston2014}. Conventional recommendations for asset management can no longer apply in this case, as they are highly dependent on wealth level, age, and Age Pension policies.

The Australian retirement system relies on three pillars - the Age Pension, the superannuation guarantee, and private savings. While the former two are mandatory, private savings are voluntary and are comprised of additional savings/assets such as voluntary superannuation contributions, private investments, and dwelling. The superannuation guarantee mandates that employers contribute a certain percentage of the employee's gross earnings to a superannuation fund, with a current contribution rate of 9.5\%. The Age Pension is the government managed safety net, which ensures a retiree meeting the age minimum and passing a means-test\footnote{Age Pension rates published by Centrelink as of September 2015. \\ (www.humanservices.gov.au/customer/services/centrelink/age-pension)} is entitled to Age Pension. In the means-test, income and assets are evaluated individually, where a certain taper rate reduces the maximum payments once income or assets surpasses set thresholds (which are subject to family status and homeownership). Income from different sources are also treated differently; financial assets are expected to generate income based on a progressive deeming rate, while income streams such as labor and annuity payments are assessed based on their nominal value. The retiree can be qualified for either full, partial, or no Age Pension. The exact amount they qualify for is determined by the smaller outcome of the income and asset tests. 

With around 80\% of the Australian population aged 65 or older being entitled to full or partial Age Pension, there is an strong need for a model that both captures the characteristics of Australian data and can explain the behavior of retirees. Empirical data indicates that 75\% of retired households are homeowners, where the home accounts for 80\% of the total wealth in middle-wealth quartiles and almost all of the wealth in the lower quartile\footnote{Estimated from data \citep{DepartmentofTreasury2010}, but consistent with \cite{Olsberg2005}}. Consumption tends to decrease with age and converges towards a consumption floor. It is argued that this decline in consumption is due to declining health of retirees, which reduces their capacity for activity \citep{Clare2014, Yogo2011}, or retirees having fewer resources reserved for consumption due to longevity risk \citep{Milevsky2011}. \cite{Higgins2011} find that the rate of decline with age is also dependent on wealth, and expenditures tend to converge towards a constant level as the retiree ages. A model that properly captures these characteristics is required to estimate the wealth needed in retirement, and to forecast Age Pension budgets and policy changes from a government perspective.

It is common to model decisions and behavior in retirement using expected utility maximization models based on dynamic programming techniques. Such models stem from the seminal work by \cite{Yaari1964, Yaari1965}, which was later extended by \cite{Samuelson1969} and \cite{Merton1969, Merton1971} who studied the problem in relation to optimal portfolio allocation. Since then, a vast amount of research has been carried out on various extensions and alternative models, where the decumulation phase of life-cycle modeling has been a topic of interest. That said, there is limited research where the Age Pension is based on a means-tested pension function. Research has mainly focused on the effect on the economy from different means-tested policies and the impact on savings (\citealp{Hubbard1995, Hurst2004, Neumark1997} to name a few).

In Australia, which is a very special case of means-testing, there have been two main approaches to model Age Pension --- expected utility maximization via dynamic stochastic programming, or dynamic general equilibrium (DGE) models. The DGE models tend to focus on how policy-changes affect welfare. Using overlapping generations DGE models, \cite{Kudrna2011b} and \cite{Tran2014} evaluate the effect of the means-tests and taper rate changes on the welfare outcome, while \cite{Kudrna2011a} examine the implication of the recent Australian pension reform extended with a housing market which is further extended in \cite{Kudrna2014}. 

While DGE models tend to analyze the effects on the economy, dynamic stochastic programming models focus on the retiree as an individual. \cite{Hulley2013} investigate the effect of the Age Pension on consumption and investments under constant relative risk aversion (CRRA) utility, while \cite{Bateman2007} compare the effect of modeling optimal consumption and investment with hyperbolic absolute risk aversion (HARA) utility against CRRA. \cite{Ding2013} captures the behavior of Australian retirees with regards to consumption, portfolio allocation, and housing in a semi-analytical HARA utility model. \cite{Iskhakov2015} models the behavior with a HARA utility maximization approach, but with the purpose of investigating how annuity purchases are affected by different preferences and scenarios. The reason HARA is preferred over CRRA in the majority of research is the presence of a consumption floor, which implies a simplified preference for risk as the consumption floor needs to be maintained. This indicates that financial advice is non-scalable with respect to wealth.

In addition to the aforementioned, the most important extensions to traditional utility maximization models from an Australian retirement perspective include housing \citep{Yogo2011, ChoSane2013}, bequest \citep{Lockwood2014, DeNardi2004, Ameriks2011} and health \citep{Yogo2011}.

A limiting factor in models is the desire for analytical solutions. For an analytical solution to exist it either constrains the number of stochastic variables allowed, or the structure and assumptions of the model. Solving the problem with realistic features therefore require numerical solutions. In addition, most research studies do not involve calibration against empirical data. \cite{Tran2014} discuss the selection of model parameters which either match Australian economy rates or are taken from other research in order to ensure that a general equilibrium steady state is achieved, but are not calibrated against data. Calibration in \cite{Ding2013} is based on mean squared error, but residuals are normalized with estimated lifetime wealth. This allows for larger errors in poor households (but improved fit for wealthier households) as the estimated pension received will be rather large in relation to actual wealth. Such normalization may affect the outcome from using said model, especially in forecasting future Age Pension budgets. \cite{Ameriks2011} calibrates a bequest model using maximum likelihood estimation, in order to investigate the bequest motives of the US data by separating precautionary savings from bequest. 

The contribution of our paper is a sequential family status model in the retirement stage that takes into consideration stochastic wealth, stochastic family status and mortality risk, and a health status proxy to allow the model to better fit empirical data. It is the first sequential model in retirement modeling to our knowledge. Our model is based on \cite{Ding2013}, but can be considered a more realistic extension of his semi-analytical model as we relax prior assumptions of deterministic investment assets and the order of means-tests phases, apply a more strict calibration model and account for mortality risk. We focus on the decumulation phase of the life cycle problem, hence from the point in time the individual retires. The model is set up as a stochastic dynamic programming problem and solved numerically via backward recursion. We calibrate the model with Australian empirical data for consumption and housing, and estimate suitable parameters via the maximum likelihood method.

The paper is structured as follows. In Section 2 we define the model to include the Age Pension means-test and present the corresponding optimal stochastic control problem. Section 3 explains the model calibration process and the assumptions imposed. Section 4 contains a discussion of the results, and concluding remarks are presented in Section 5. 

\section{Model specification}
We assume that the agent's goal is to maximize the expected value of utility that is associated with consumption, housing, and bequest. Utility is measured by time-separable additive functions based on commonly used HARA utility functions. 

Let ($\Omega$, $\mathcal{F}$, $\{ \mathcal{F}_t
\}_{t \ge 0}$, $\mathbb{P}$) be a filtered complete probability space
and $\mathcal{F}_t$ represents the information available before time $t$. We assume that all the processes introduced below are well defined and adapted to $\{\mathcal{F}_t\}_{t \ge 0}$. Denote the value of liquid financial assets as random variable $W_t$ and family status as random variable $G_t$ at the agent's anniversary dates $t=t_0,  t_0+1,...,T$, where $t_0$ is the retirement age and $T$ is the maximum age of the agent beyond which survival is deemed impossible. Realizations of $W_t$ and $G_t$ are denoted as $w_t$ and $g_t$ respectively. The utility received at times $t$ is subject to the agent decision (control) variables $\alpha_t$ (proportion drawdown of liquid assets), $\delta_t$ (proportion liquid assets allocated to risky assets) and the decision variable $H$ (wealth allocated to total housing only at time $t=t_0$)\footnote{Note that $H$ represents the price of the house at $t_0$, and not additional wealth invested into housing. If the retiree currently is a homeowner, then the difference between $H$ and current house value represent the change in housing suggested.}. Given a current state $x_t=(w_t,g_t)$ we can define a decision rule $\pi_t(x_t) = (\alpha_t(x_t),\delta_t(x_t))$ which is the action at time $t$ and $x_t$ is the value of the state variables before the action. Then a sequence (policy) of decision rules is given by $\pi = (\pi_{t_0}, \pi_{t_0+1}, ..., \pi_{T-1})$ for $t=t_0, t_0+1, ..., T-1$. 

The optimization problem is set to begin at the time of retirement $t=t_0$ when all available wealth is placed in an allocated pension account, and the agent decides how much wealth to allocate into a family home. Taxes after $t_0$ are not considered since earnings in an allocated pension account are tax-free. At the start of each year, $t$, the agent makes a decision of how much to withdraw from the account, and receives a means-tested Age Pension $P_t$. The remaining wealth is placed in a stochastic risky portfolio $S_t$ and risk free cash account until $t+1$, where the agent faces the decision of what proportion $\delta_t$ to allocate in risky assets. The wealth process and consumption are then given as

\begin{equation}
\label{eq:transition}
W_{t+1} = \left[ W_t - \alpha_t W_t \right] \left[\delta_t e^{Z_{t+1}} + (1-\delta_t) e^{r_t} \right], \quad W_{t_0} = \mathsf{W} - H,
\end{equation}
\begin{equation}
\label{eq:consconstraint}
C_t = \alpha_t W_t + P_t,
\end{equation}
where $W_t \ge 0$ is the liquid assets at time $t$ before withdrawal. The initial liquid assets $W_{t_0}$ are the remaining wealth after housing allocation at retirement where $\mathsf{W}$ is the total wealth at time $t=t_0$. The housing allocation is constrained with $H \in \{0, [H_L, \mathsf{W}]\}$, where $H_L \ge 0$ is a lower bound of housing, hence wealth can never be negative. We assume that risky asset $S_t$ follows a geometric Brownian motion such that the real log returns of the risky asset $Z_{t+1}=\ln (S_{t+1}/S_t)$, $t=0,1,\ldots$ are independent and identically distributed from a Normal distribution $\mathcal{N}(\mu-\widetilde{r},\sigma)$ with the mean $\mu - \widetilde{r}$ and variance $\sigma^2$, where $\widetilde{r}$ is the inflation rate. The real risk free rate $r_t$ (adjusted for inflation) is time dependent but deterministic. The constraints for consumption, where $C_t$ is the consumption and $P_t$ is the Age Pension, show that consumption equals the sum of received Age Pension and drawdown of wealth for the current period. By defining the model in real terms (adjusted for inflation) we can avoid dealing with inflation in consumption, Age Pension and consumption floor to better capture the characteristics of Australian retirees. 

The model operates on a household level, where we separate between couple and single retiree households due to the Age Pension treating couples as a single entity. The agents face the risk of family status transitions due to death each period, with the possible states defined as

\begin{equation}
G_t \in \mathcal{G} = \{\Delta, 0, 1, 2\},
\end{equation}
whether the agent is already dead at time $t$ ($\Delta$), died during $(t-1, t]$ ($0$), alive at time $t$ in a single household ($1$) or alive in a couple household ($2$), subject to survival probabilities. An agent can therefore start off at time $t=t_0$ as either a couple or single household. In case of a couple household, there is a risk each time period that one of the spouses passes away, in which case it is treated as a single household model for the remaining years. $Z_t$ and $G_t$ are assumed to be independent, hence a large investment loss does not affect the death probabilities (though one can argue that it might affect i.e. the ability to pay hospital bills or life quality, which in turn affects death probabilities). 

Each period the agent receives utility based on the current state of family status $G_t$:
\begin{equation}
\label{RewardFunction}
 R_{t}(W_t,G_t,\alpha_t,H) = \left\{ \begin{array}{ll}
         U_C(C_t,G_t,t) + U_H(H,G_t), & \mbox{if $G_t = 1,2$},\\
         U_B(W_t), & \mbox{if $G_t = 0$},\\
         0, & \mbox{if $G_t = \Delta.$}\end{array} \right. 
\end{equation} 
That is if the agent is alive he receives reward based on consumption $U_C$ and housing $U_H$, if he died during the year the reward comes from the bequest $U_B$ and if he is dead there is no reward. Note that the reward received when the agent is alive depends on whether the state is a couple or single household due to differing utility parameters and Age Pension thresholds. The terminal reward function at $t=T$ is given as

\begin{equation}
\label{TerminalRewardFunction}
\widetilde{R}(W_T,G_T) = \left\{ \begin{array}{ll}
         U_B(W_T), & \mbox{if $G_T \geq 0,$}\\
         0, & \mbox{if $G_T = \Delta.$}\end{array} \right. 
\end{equation} 
         
The current model is based on \cite{Ding2013}, but with the addition of sequential family status, mortality risk and risky asset, and a different health proxy. The retiree wants to find the policy that maximizes expected utility with respect to the decision for consumption, investment and housing. This is defined as a stochastic control problem 

\begin{equation}
\label{eq:FinalValueFunction}
\widetilde{V} := \underset{H}{\max} \left[ \underset{\pi}{\sup} \: \mathbb{E}^\pi_{t_0} \left[\beta_{t_0,T} \widetilde{R}(W_T,G_T) + \sum_{t={t_0}}^{T-1} \beta_{t_0,t} R_{t}(W_t,G_t,\alpha_t,H)\right] \right],
\end{equation}
that can be solved with dynamic programming by using backward induction of the Bellman equation. $\mathbb{E}^\pi_{t_0}[\cdot]$ is the expectation conditional on information at time $t=t_0$ if we use policy $\pi$ up to $t=T-1$. The policy $\pi$ contains the control variables for each time period and $\beta_{t,t'}$ is the discounting from $t$ to $t'$.	

The subjective discount rate $\beta_{t,t'}$ is a proxy for personal impatience between time $t$ and $t'$, and set in relation to the real interest rate so that 

\begin{equation}
\label{DiscountFactor}
\beta_{t,t'} = e^{-\sum_{i=t}^{t'}r_i}.
\end{equation}
This assumption suggests that optimal consumption rates would be constant over time for HARA utility in the absence of mortality risk and risky investments \footnote{This can easily be shown with simple calculus.}.

\subsection{Consumption preferences}
We assume that the HARA utility comes from consumption exceeding the consumption floor, weighted with a time dependent health status proxy. The utility function for consumption is defined as

\begin{equation}
\label{eq:consumption}
U_C (C_t,G_t,t) = \frac{1}{\psi^{t-t_0} \gamma_d} \left(\frac{C_t - \overline{c}_d}{\zeta_d} \right)^{\gamma_d}, \quad d = \left\{ \begin{array}{ll}
         \mathrm{C}, & \mbox{if $G_t = 2 \quad \text{(couple),}$}\\
         \mathrm{S}, & \mbox{if $G_t = 1 \quad \text{(single),}$}\end{array} \right.
\end{equation}
where $\gamma_d \in (-\infty,0)$ is the risk aversion and $\overline{c}_d$ is the consumption floor parameters. The scaling factor $\zeta_d$ normalizes the utility a couple receives in relation to a single household\footnote{If single and couple households had the same risk aversion and no scaling factor was used, the solution would suggest similar consumption for both. This effect comes from the consumption smoothing properties of life cycle models, hence needs to be adjusted as a couple household's utility is for two people. It would otherwise cause problems in the calibration stage.}. The utility parameters $\gamma_d$, $\overline{c}_d$ and $\zeta_d$ are subject to family state $G_t$, hence will have different values for couple and single households. Note that the consumption for any period is based on the pension $P_t$ received in the same period, and the drawdown $\alpha_t$ from the liquid assets $W_t$ as given by equation (\ref{eq:consconstraint}). The proportion of wealth drawn down can be positive or negative, with a negative value indicating that part of the pension received is saved for future consumption. As consumption tends to converge towards a consumption floor as the retiree ages despite wealth status, we define a health proxy to control the slope of the decline. Let $\psi \in [1,\infty)$ be the utility parameter for the slope where the difference between current time $t$ and time of retirement $t_0$ determines the power of the parameter. This allows the initial time $t = t_0$ to not be affected by a health proxy, and as the retiree ages the slope of the proxy decreases to allow consumption to converge over age and wealth groups. This decreasing convex health proxy has a better fit to empirical data, compared with survival probabilities as used in \cite{Ding2013} which is decreasing but concave.

\subsection{Bequest preferences}

We adopt the bequest function in \cite{Lockwood2014} which is a re-parameterized version of \cite{DeNardi2004}, as the parameters are slightly more intuitive. The utility is received from \emph{luxury} bequest, hence the home is not included in the bequest\footnote{As housing is both a necessity and bequest it cannot be treated the same as other bequest \citep{Ding2014}, since the intentional bequest component is difficult to separate. By using luxury bequest the model can better explain inequalities between wealth percentiles as wealthier retirees tend to bequeath a larger proportion of their assets. Previous research finds support for luxury bequest such as \cite{DeNardi2010} and \cite{Lockwood2014}.} and is not considered for such a purpose at the time of purchase. The utility function is then defined as

\begin{equation}
U_B(W_t) = \left(\frac{\theta}{1-\theta}\right)^{1-\gamma_\mathrm{S}} \frac{\left(\frac{\theta}{1-\theta} a+W_t\right)^{\gamma_\mathrm{S}}}{\gamma_\mathrm{S}},
\end{equation}
where $W_t$ is the liquid assets available for bequest, $\gamma_\mathrm{S}$ the risk aversion parameters of bequest utility (which is considered to be the same as consumption risk aversion for singles, since a couple is expected to become a single household before bequeathing assets)\footnote{In case couple households have a different risk aversion towards bequest it will be absorbed by adjusting the ratio of single and couple risk aversion.} and the two parameters $\theta \in \left[0,1\right)$ and $a \in \mathbb{R}^+$. The threshold for luxury bequest, $a$, is the threshold up to where the retiree leaves no bequest\footnote{There is strong empirical evidence that wealthy retirees leave a larger proportion of their wealth as bequest compared with less wealthy (\citet{Ameriks2011}, \citet{Hurd2002}, \citet{Ding2013}).}. If $a=0$ then consumption and bequest are homothetic. The degree of altruism, $\theta$, controls the preference of bequest over consumption. Low values of $\theta$ reflect that retirees prefer consumption over bequest, while higher values decrease the marginal utility of bequest. As $\theta \rightarrow 1$ the bequest motive approaches a linear function with a constant marginal utility of $\frac{dU_B}{dW_t} = a^{\gamma-1}$.

\subsection{Housing preferences}
Housing differs from other assets in that it provides a flow of services in terms of the preference (utility) of owning a house compared with renting, in addition to its residual value. We apply the assumption that the utility is linked to the house value as in \cite{Ding2013} and \cite{ChoSane2013}. The utility from owning a home is defined as
\begin{equation}
U_H (H,G_t) = \frac{1}{\gamma_\mathrm{H}} \left(\frac{\lambda_d H}{\zeta_d} \right)^{\gamma_\mathrm{H}},
\end{equation}
where $\gamma_\mathrm{H}$ is the the risk aversion parameter for housing (allowed to be different from risk aversion for consumption and bequest), $\zeta_d$ is the same scaling factor as in equation (\ref{eq:consumption}), $H > 0$ is the market value of the family home at time of purchase $t_0$ and $\lambda_d \in (0,1]$ is the preference of housing defined as a proportion of the market value. 

Note the house value $H$ is not indexed with time. The retiree decides how much of his wealth to allocate to housing at the time of retirement $t_0$, which remains constant afterwards. We do not consider the house to be a liquid asset, rather a proxy for the utility received by being a homeowner. Our assumptions reflect the housing behavior of Australian retirees. Most Australian households do not convert housing assets to liquid assets in order to cover expenses in retirement, with the exception of certain events (such as death of a spouse or move to an Aged Care facility) \citep{Olsberg2005}. Wealthier retirees prefer to invest more in the family home as wealth increases, but with a decreasing marginal utility since the percentage allocation decreases consistent with the utility model used.

\subsection{Age Pension function}

We assume that all liquid assets are converted into an allocated pension account at the time of retirement, as for the year of our data sample over 90\% of income came from allocated pensions. This type of account has the advantage that earnings on assets are tax free and allows for an yearly income test deduction. The Age Pension received is modeled with respect to the current liquid assets. The account value is used in the asset test, and the drawdown from this account is the income generated for the income test. Based on these assumptions, the Age Pension function can be defined as:

\begin{equation}
P_t := f(\alpha_t, W_t, t) = \max \left[0, \min \left[P^{d}_\mathrm{max}, \min \left[P_{\mathrm{A}}(W_t), P_{\mathrm{I}}(\alpha_t W_t,t)\right] \right] \right],
\end{equation}
where

\begin{equation}
P_{\mathrm{A}}(W_t) = P^d_\mathrm{max} - (W_t-L^{d,h}_{\mathrm{A}})\varpi^d_{\mathrm{A}},
\end{equation}
\begin{equation}
P_{\mathrm{I}}(\alpha_t W_t,t) = P^d_\mathrm{max} - (\alpha_t W_t - M(t) - L^{d}_{\mathrm{I}})\varpi^d_{\mathrm{I}}.
\end{equation}
Here, $P^d_{\mathrm{max}}$ is the full Age Pension, $L^d$ is the threshold for the asset and income test respectively (as indicated by subscript) and $\varpi^d$ the taper rate for assets/income exceeding the thresholds, and superscript $d$ is a categorical index indicating couple or single household status as defined in equation (\ref{eq:consumption}). The variables are subject to whether it is a single or couple household, and the threshold for the asset test is also subject to whether the household is a homeowner or not $h=\{0,1\}$ (see Table \ref{table:PensionRates} for parameterization of the function). The function $M(t)$ is an income test deduction set when the wealth is converted into an allocated pension account\footnote{Income streams commenced after 1\textsuperscript{st} January 2015 may no longer include an income test deduction. Income is now determined by a deeming rate applied to assets. See \cite{fahcsia2016} regarding this function.}, defined as:

\begin{equation}
M(t) = \frac{W_{t_0}}{e_{t_0}}(1+\widetilde{r})^{t_0-t},
\end{equation}
where $e_{t_0}$ is the lifetime expected at age $t_0$ and $\widetilde{r}$ the inflation. As the model is defined in real terms, the future income test deductions must discount inflation.

\subsection{Solution as a stochastic control problem}
\label{StochControlProblem}
To solve the optimal stochastic problem of maximizing expected utility with respect to the decision policy, given by equation (\ref{eq:FinalValueFunction}), we follow the theory and notation specified in \cite{Bauerle2011} to define the dynamic programming problem.

The starting point is the basic model where the wealth $W_t$ and family status $G_t$ are stochastic, and the terminal time $T$ (time beyond which survival is deemed to be impossible) is fixed. The problem is defined as follows
\begin{itemize}
\item Denote a state vector as $X_t=(W_t,G_t) \in \mathcal{W} \times \mathcal{G}$ where $W_t \in \mathcal{W} = \mathbb{R}^+$ denotes the current level of wealth and $G_t \in \mathcal{G} = \{\Delta, 0, 1, 2\}$ whether the agent is dead, died this period, alive in a  single household or alive in a couple household. The stages are sequential hence an agent that starts out as a couple becomes single when one spouse dies.

\item Denote an action space of $(\alpha_t, \delta_t) \in \mathcal{A} = (-\infty,1]  \times [0,1]$ for $t=t_0, ..., T-1$ where $\alpha_t \in (-\infty,1]$ denotes the proportion of wealth consumed and $\delta_t \in [0,1]$ is the percentage of wealth allocated in the risky asset. The upper boundary of 1 indicates that drawdown cannot be larger than our total wealth, hence borrowing is not allowed. Negative values for drawdown are allowed however as they represent savings from Age Pension into wealth.

\item Denote an admissible space of state-action combination as $D_t(x_t) = \{\pi_t(x_t) \in \mathcal{A} \mid \alpha_t \ge \frac{\overline{c}_d-P_t}{w_t}\}$ which contains the possible actions for the current state, and indicates that withdrawals must be large enough to cover the consumption floor, net of Age Pension received.

\item The transition function $T_t(W_t,\alpha_t,\delta_t,z_{t+1}) := W_{t+1} = W_t (1-\alpha_t) \times (\delta_t e^{z_{t+1}} + (1-\delta_t) e^{r_t})$ where $z_{t+1}$ is the realization of the log return on the stochastic investment portfolio over $(t,t+1]$. We assume the agent is small and cannot influence asset price.

\item Denote the stochastic transitional kernel as $Q_t(dx'|x,\pi_t(x))$ which represents the probability of reaching a state in $dx' = (dw_{t+1},g_{t+1})$ at time $t+1$ if action $\pi_t(x)$ is applied in state $x$ at time $t$. Since the transition function is based on the stochastic risky return $Z_{t+1}$, which is Markovian, the transition probability for wealth $W_{t+1}$ is determined by the distribution of the risky return, where $Z_{t+1} \overset{i.i.d}{\sim} \mathcal{N}(\mu-\widetilde{r},\sigma^2)$ with the probability density function denoted as $f_\mathcal{N}(z)$. Let $q(g_{t+1},g_t)$ denote $\Pr[G_{t+1}=g_{t+1} \mid G_t = g_t]$. Since both state variables depend on exogenous and independent probabilities, we have
\begin{equation}
\begin{aligned}
\label{StochasticKernel}
&Q_t(dx'|x,\pi_t(x),\pi)
\\ & \quad = \Pr [W_{t+1} \in dw_{t+1}, G_{t+1}=g_{t+1} \mid X_t = x_t] 
\\ & \quad = \Pr [T_t(W_t,\alpha_t,\delta_t,Z_{t+1}) \in dw_{t+1}, G_{t+1}=g_{t+1} \mid W_t = w_t, G_t = g_t]
\\ & \quad = \Pr [T_t(W_t,\alpha_t,\delta_t,Z_{t+1}) \in dw_{t+1} \mid W_t = w_t ]\times q(g_{t+1},g_t).
\end{aligned}
\end{equation}
The probabilities for family status are defined as
\begin{equation}
\begin{aligned}
&q(2,2) = p^\mathrm{C}_t, \quad q(1,2) = 1-p^\mathrm{C}_t, \\
&q(1,1) = p^\mathrm{S}_t, \quad q(0,1) = 1-p^\mathrm{S}_t, \\
&q(\Delta,0) = q(\Delta,\Delta) = 1, \\
\end{aligned}
\end{equation}
where $p^\mathrm{C}_t$ is the probability of surviving one more year as couples or $p^\mathrm{S}_t$ as singles. All other transition probabilities for family status have probability $0$.

\item The reward function depends on the $G_t$ state as defined in equation (\ref{RewardFunction}). If the agent is alive he receives a reward based on consumption, if he died during the year the reward comes from bequest, and if he is dead there is no reward. Note that the reward when alive depends on the Age Pension received and the consumption floor, which differs for couples and singles.

\item The terminal reward function is defined in equation (\ref{TerminalRewardFunction}).

\item The discount factor is defined in equation (\ref{DiscountFactor}) earlier, with  $\beta_{t,t+1} \in (0,1]$.
\end{itemize}

The optimal value function can now be stated as in equation (\ref{ValueFunction}) at starting time $t_0$. The value function can be split into the stochastic control problem with respect to $\alpha_t$ and $\delta_t$, and the decision problem for $H$ at $t=t_0$ due to the time-separable additive preferences of the utility functions. A solution for the stochastic control problem is given by a backward recursion Bellman equation
\begin{equation}
\label{ValueFunction}
V_t(W_t,G_t) = \underset{\pi_t(x_t) \in D_t(x_t)}{\sup} \left\{R_{t}(W_t,G_t,\alpha_t,H) + \beta_{t,t+1} \, \mathbb{E}^\pi_{t} \left[ V_{t+1}(W_{t+1},G_{t+1}) \mid W_t, G_t \right] \right\},
\end{equation}
where $\mathbb{E}^\pi_t[\cdot]$ is calculated using the stochastic transition kernel $Q(\cdot)$ given in equation (\ref{StochasticKernel})
\begin{equation}
\mathbb{E}^\pi_{t} \left[ V_{t+1}(W_{t+1},G_{t+1}) \mid W_t, G_t \right] = \sum_{g_{t+1} \in \mathcal{G}} \int^{\infty}_{-\infty} V_{t+1}(W_{t+1},G_{t+1})f_{\mathcal{N}}(z_{t+1})dz \times q(g_{t+1},g_t).
\end{equation}
It is shown in Appendix \ref{AppendixHousing} that the problem can be simplified if we introduce an alternative reward function without the housing utility
\begin{equation}
 \overline{R}_{t}(W_t,G_t,\alpha_t) = \left\{ \begin{array}{ll}
         U_C(C_{t},G_t,t), & \mbox{if $G_t = 1,2,$}\\
         U_B(W_t), & \mbox{if $G_t = 0,$}\\
         0, & \mbox{if $G_t = \Delta,$}\end{array} \right. 
\end{equation}
where the corresponding Bellman equation becomes
\begin{equation}
\label{VBar}
\begin{aligned}
\overline{V}_t(W_t,G_t) = & \underset{\pi_t(x_t) \in D_t(x_t)}{\sup} \left\{\overline{R}_{t}(W_t,G_t,\alpha_t) + \beta_{t,t+1} \, \mathbb{E}^\pi_{t} \left[ \overline{V}_{t+1}(W_{t+1},G_{t+1}) \mid W_t, G_t \right] \right\} \\
= & \underset{\pi_t(x_t) \in D_t(x_t)}{\sup} \{\overline{R}_{t}(W_t,G_t,\alpha_t)\\
& + \sum_{g_{t+1} \in \mathcal{G}} \int^{\infty}_{-\infty} \overline{V}_{t+1}(T_t(W_t,\alpha_t,\delta_t,z_{t+1}),g_{t+1})f_{\mathcal{N}}(z_{t+1})dz \times q(g_{t+1},g_t) \},
\end{aligned}
\end{equation}
and the terminal condition at time $t=T$ is 
\begin{equation}
\overline{V}_T(W_T,G_T) = \widetilde{R}_{T}(W_T,G_T).
\end{equation}
This allows us to avoid the house value being a state variable, effectively eliminating one dimension and allowing the problem to be solved as
\begin{equation}
\label{eq:housingVSwealth}
\widetilde{V} = \underset{H\in[0,\mathsf{W}]}{\max} \left[ \overline{H}_{t_0}(H,G_{t_0}) + \overline{V}_{t_0}(\mathsf{W}-H, G_{t_0}) \right],
\end{equation}
where $\overline{H}_t(H,G_t)$ is the summation of housing utility weighted with survival probabilities at time $t$ given as
\begin{equation}
\overline{H}_t(H,G_t) = \left\{ \begin{array}{lll}
		\sum^{T-1}_{i=t} & \beta_{t,i} [ \,\,_{t}p^{\mathrm{C}}_{i} \, U_H(H,G_t=2) & \mbox{if $G_t = 2,$}\\
		& + \,_{t}p^{\mathrm{C}}_{i-1} (1-p^{\mathrm{C}}_{i-1})\overline{H}_t(H,G_t=1)], & \\
         \sum^{T-1}_{i=t} & \beta_{t,i} \,\,_{t}p^{\mathrm{S}}_i \, U_H(H,G_t=1), & \mbox{if $G_t = 1,$}\\
         0, & & \mbox{if $G_t = 0, \Delta,$}\end{array} \right. 
\end{equation}
where $_tp^d_{t'}$ denotes the probability of surviving from year $t$ to year $t'$ for singles ($d=\mathrm{S}$) or couples ($d = \mathrm{C}$).
The validity of the problem setup and existence of optimal policies is implied by the integrability assumption and structure assumption\footnote{\cite{Bauerle2011} shows the integrability assumption holds if the reward and terminal function are bounded from above and satisfies $\mathbb{E} Z_n < \infty$ for all $n$ where $Z_n$ is the disturbance term. A power utility function with $\gamma < 0$ has an upper bound of 0, and a log-normal random variable has a finite expected value, hence the integrability assumption is satisfied.}.

\subsection{Numerical implementation}
The model is solved numerically. By discretizing the wealth state on a grid of log-equidistant grid points $W_0,...,W_k$ for each year $t=t_0,...,T$ we solve the Bellman equation for value function $\overline{V}$ (equation (\ref{VBar})) recursively with backward induction. The lower bound of the grid $W_0$ is set to \$1 as the utility for \$0 does not exist and the upper bound $W_k$ is determined by the maximum wealth in the dataset and the risky asset. We use $W_k = \hat{W}_{\mathrm{max}} e^{(T-t_0)\mu + 5\sqrt{T-t_0}\sigma}$ where $\hat{W}_\mathrm{max}$ is the largest wealth sample to find a conservative upper bound. This means that no extrapolation is needed when integrating risky returns and so values close to the upper bound have no material effect on the range $[W_0,\hat{W}_{\mathrm{max}}]$ actually used in the solution. For each grid point in the wealth state we find an optimal drawdown proportion $\alpha_t$ and risky asset allocation $\delta_t$ with a 2-dimensional optimization. 

The value function $\overline{V}$ is interpolated between grid points based on the shape preserving Piecewise Cubic Hermite Interpolation Polynomial (PCHIP) method, which preserves the monotonicity and concavity of the value function \citep{Kahaner1988}. The need to interpolate arises from the integration of the stochastic return. Since the value function is only available at predefined grid points, any values in between needs to be interpolated. If traditional cubic splines were to be used there is a high chance of the function overshooting a point, hence the solution could return a local rather than global maximum. Linear interpolation requires a much higher grid density at lower values due to the steep derivative in these regions, hence PCHIP is preferred. In general, given a function $f(x)$ and state $x$ between two grid points $x_k \le x \le x_{k+1}$ the interpolant $P(x)$ of $f(x)$ for the $k$th interval is calculated as
\begin{equation}
\begin{aligned}
P(x)= & \frac{3h_k(x-x_k)^2 - 2(x-x_k)^3}{h_k^3} f(x_{k+1})+\frac{h_k^3-3h_k(x-x_k)^2+2(x-x_k)^3}{h_k^3} f(x_k) \\ & + \frac{(x-x_k)^2(x-x_k-h_x)}{h_k^2} d_{k+1}+\frac{(x-x_k)(x-x_k-h_k)^2}{h_k^2}d_k,
\end{aligned}
\end{equation}
where $h_k = x_{k+1}-x_k$ and the slope $d$ of the interpolant depends on the first divided difference $\Delta_k = (f(x_{k+1})-f(x_k))/h_k$. If $\Delta_k$ and $\Delta_{k-1}$ are of opposite polarity then $d_k = 0$, otherwise $d_k$ is given by
\begin{equation}
\frac{3h_k+3_{k-1}}{d_k} = \frac{2h_k+h_{k-1}}{\Delta_{k-1}} + \frac{h_k+2h_{k-1}}{\Delta_k}.
\end{equation}
In addition to this, the conditions $P(x_k)=f(x_k)$, $P(x_{k+1})=f(x_{k+1})$, $P'(x_k)=d_k$ and $P'(x_{k+1})=d_{k+1}$ must hold true.

The expectation with respect to the stochastic return in equation (\ref{VBar}) is calculated with Gauss-Hermite quadrature 
\begin{equation}
\int^\infty_{-\infty} e^{-x^2}f(x)dx\approx \sum^M_{i=1} w(x_i) f(x_i),
\end{equation}
where $w(x_i)$ is the weight and $x_i$ is the node at which to evaluate the value function, which gives an exact result if $f(x)$ is any polynomial up to the order $2M-1$. For details on how to find the weights and nodes see a textbook on numerical integration, such as \cite{Kahaner1988}. The expectation is then calculated as 
\begin{equation}
\begin{aligned}
\int_{-\infty}^{\infty} & \overline{V}_{t+1}(T_t(W_t,\alpha_t,\delta_t,z),G_{t+1})f_{\mathcal{N}}(z) \; dz \\ 
& = \int_{-\infty}^{\infty}\frac{e^{-x^2}}{\sqrt{\pi}} \overline{V}_{t+1}(T_t(W_t,\alpha_t,\delta_t,\sqrt{2} \sigma x + \mu),G_{t+1}) \; dx \\
& \approx \sum_{i=1}^{M} \frac{w(x_i)}{\sqrt{\pi}} \; \overline{V}_{t+1}(T_t(W_t,\alpha_t,\delta_t,\sqrt{2} \sigma x_i + \mu),G_{t+1}),
\end{aligned}
\end{equation}
using $M=5$ nodes\footnote{Solving the Bellman equation with 5, 10 or 25 nodes resulted in negligible differences, hence 5 nodes were chosen to reduce calculation time.}. To speed up calculations, a temporary vector is created each year where the expectation of the value function is calculated for each grid point. By doing this prior to finding the optimal control we avoid repeating the numerical integration during the optimization as it is now enough to interpolate. 

Finally, once the backward induction has reached $t=t_0$ the initial wealth $W_0$ can be determined by optimizing the allocation between housing and liquid assets in equation (\ref{eq:housingVSwealth}). The optimal path for a retiree can then be derived by following the optimal drawdown and risky allocation from the policy that corresponds with our wealth grid point for each time $t$ and keep repeating until the terminal condition at $t=T$.

We performed tests with additional number of nodes for the Gauss-Hermite quadrature and larger bounds for $W_k$ to ensure the accuracy of our numerical solution. In addition to this a forward Monte Carlo simulation with random policies was generated to verify optimality of the solution.

\section{Calibration}

We calibrate the model using a similar approach and the same data as \cite{Ding2013}, but with maximum likelihood estimation instead of mean squared errors in order to estimate utility parameters.

The calibration is rather computationally expensive. Each time new parameters are suggested the model needs to be solved 4 times (each combination of single/couple households and homeowner/non-homeowner), and the model output needs to be generated from the wealth, age and homeowner data given in each sample. The model output is then compared with the sample data for consumption and housing to estimate the fit of the model as in Section \ref{sec:calibrationmodel} before the next iteration begin.

The calibration is carried out in two steps. First, a suitable starting point is identified by searching globally. Each utility parameter is assigned a realistic range of values where 3 different values are selected, hence 1,000 iterations are carried out. Once a starting point is identified, the parameters are optimized further using the Nelder-Mead Simplex algorithm until further improvements of the log likelihood function are negligible.

\subsection{Dataset}
\label{sec:dataset}
For the dataset we use the Household Expenditure Survey (HES) 2009-2010 and the Survey of Income and Household (SIH) 2009-2010 from Australian Bureau of Statistics \citep{Statistics2011}. This dataset has a limitation in that data is only collected for private households, hence retirees within assisted care facilities are excluded. Out-of-pocket health expenses are included, but any costs associated with private facilities are not. Furthermore, we do not generalize between households with or without dependents and treat them as one group, even if consumption in reality might differ.

The samples are filtered by labor status ('not in work force') and the age requirement for Age Pension to find eligible retirees, where the age of a couple household is based on the youngest spouse. The data is then aggregated for each sample to reflect total expenditure (excluding mortgage payments), family home value, and wealth. In order to clean the data from possible reporting errors, we remove samples that received no Age Pension despite being entitled to a material portion, as well as any samples with expenditure less than \$3,000 per year or larger than assets available (liquid assets and Age Pension). Since the model is restricted to no borrowing any entries with negative wealth are removed as well. This resulted in 2,017 samples for couple households and 2,038 samples for single households.

\subsection{Assumptions}

In order to ensure a realistic calibration, some assumptions and constraints are needed. We impose the following:

\begin{itemize}
\item[-] For samples with age older than the entitlement age, we assume that the wealth at retirement equals the wealth of the sample, in order to calculate the income test deduction $M(t)$.
\item[-] We assume that households are aware of their life expectancy hence can take this into consideration for decisions.
\item[-] A potential home owner is required to have sufficient funds for an initial down payment on a home. As the first wealth quartile in the dataset are unlikely to be home owners, a lower threshold for housing is set to \$30,000 to make this consistent with the data. A retiree with wealth below this level can therefore not be a homeowner.
\end{itemize}

The following constraints have been imposed on the utility parameters to ensure meaningful variables, rather than over-fitting the model to the sample data. That said, none of the constraints were binding for the parameters once the calibration finished.

\begin{description}
\item $\overline{c}_d \in [0,P_{\mathrm{max}}]$, which ensures that the consumption floor does not violate budget constraints for poor households. In other words, the necessary spending cannot be higher than future income in case of no accumulated wealth.
\item $\gamma_d < 0$, since the utility function is discontinuous at $\gamma_d = 0$ and positive values indicate risk seeking behavior.
\item $\theta \ge 0$, preference of bequest over consumption must be positive as we do not allow negative bequest.
\item $a \ge 0$, threshold for luxury bequest must be positive.
\item $\lambda_d \ge 0$, utility parameter from being a homeowner cannot be negative, as otherwise the optimization might suggest that selling a house the retiree does not own while still receiving utility is optimal.
\end{description}

\subsubsection{Survival probabilities}
The sequential model introduces two extra dimensions to the calibration; the age of the second person in a couple household, and the age difference between the spouses. In addition to this, survival probabilities differ between females and males, which would require additional parallel solutions to the model. If the age difference in couples were considered this would be very computationally expensive and unrealistic for calibration. We therefore generalize survival probabilities for single and couple households into a single unisex dimension.

In order to estimate the unisex survival probability, the ratio of male to females alive (estimated from the cumulative probability to be alive) is used as weights. The estimated ratios match the empirical data proportions of males and females alive at any age $t$ in \cite{Statistics2011}, and can be used as a proxy for survival probabilities to avoid the gender variable. The unisex probability of surviving one more year at age $t$ is defined as
\begin{equation}
p^\mathrm{S}_t = 1-\frac{q^\mathrm{M}_t \times \,_{t}p^\mathrm{M}_0 + q^\mathrm{F}_t \times \,_{t}p^\mathrm{F}_0}{\,_{t}p^\mathrm{M}_0 + \,_{t}p^\mathrm{F}_0},
\end{equation}
where superscript indicates a single unisex household (S), male (M) or female (F) probabilities, and $\,_{t}p_0$ is the probability of surviving from birth (year 0) to age $t$. The actual mortality probabilities are taken from Life Tables published in \cite{ABSMortality2012}.

The assumptions for couple households are different since we already know that both spouses are alive, hence no weighting of male to females ratios are necessary. The events of independent deaths of each spouse are non-mutually exclusive, however, we treat them as if they were mutually exclusive to follow the model assumptions. We do not expect both spouses to die the same year due to the low probability of this occurring and the effect it would have on the solution is minimal. The probability of surviving one more year as a couple at time $t$ is therefore defined as
\begin{equation}
p^\mathrm{C}_t = 1- \left( q^\mathrm{M}_t + q^\mathrm{F}_t \right).
\end{equation}

\subsubsection{Portfolio composition and returns}
The expected return and volatility have an important effect on the model calibration. To make the wealth process as realistic as possible, we estimate a typical portfolio composition of Self-Managed Super Fund (SMSF) accounts and then use this typical composition with longer term financial data to find portfolio return. This way we can use the actual portfolio returns in the calibration, rather than returns based on the optimal allocation control parameter (which most likely will not be a correct representation for the average retiree).

In order to estimate the typical portfolio composition we use SMSF data for each financial year from 2008 to 2014\footnote{Data contains individual account balances, earnings and drawdowns each year for Self-Managed Super Fund accounts, taken from a dataset provided by Australian Taxation Office to CSIRO-Monash superannuation research cluster (not publicly available).}, from which we calculate actual investment returns on SMSF accounts. The average operating expense ratio reported by ATO for the same period as the SMSF data is 0.83\%\footnote{Data is taken from 'Australian Taxation Office Reports SMSFs: A statistical overview' for the years 2008-09, 2009-10, 2010-11, 2011-12 and 2012-13}. The portfolio is assumed to be based on a set of risky assets  approximated with S\&P/ASX 200 Total Return which includes dividends, and a risk free asset approximated with the deposit interest rate. The portfolio weight (proportion risky assets) $\delta$ is then estimated with least square regression by regressing the average SMSF account returns for each year against the returns of S\&P/ASX 200 Total Return and the deposite rate. We find $\delta$ to be 43.7\% with a significance level of 1\%. This $\delta$ is used as a proxy for risky asset allocation during calibration of the model.

To estimate the long term returns we take the 20 year average log-returns prior to 2010 of S\&P/ASX 200 Total Return. The returns are then adjusted to real returns by deducting inflation ($\widetilde{r} = 2.9\%$) and for operating expenses. The final estimates give $r_t = 0.005$ and $Z_t \sim \mathcal{N}(0.056, 0.133)$ which are used in equation (\ref{eq:transition}).

\subsection{Parameters}
The parameters for the Age Pension are shown in Table \ref{table:PensionRates}, and taken for the year 2010 to match the data. A retiree is eligible for Age Pension at age 65 (male) or 63 (female). We set the scaling factor for couple households to $\zeta = 1.3$, which is in line with the results in \cite{Fernandez-Villaverde2007} who review research in controlling for family size and the resulting economy of scale. In addition to this, we set $T=100$.  

\begin{table}[h]
\centering
\caption{Age Pension rates published by Centrelink as at January 2010}
\label{table:PensionRates}
\begin{tabular}{l c c}
\hline
\multicolumn{2}{r}{Single} & Couple\\
\hline
Full Age Pension Rate ($P^d_\mathrm{max}$)& \$17,456 & \$26,099\\
\hline
Income Test\\
Threshold ($L^{d}_\mathrm{I}$) & \$3,692 & \$6,448\\
Rate of Reduction ($\varpi^d_{\mathrm{I}}$) & \$0.5 & \$0.5\\
\hline
Asset Test\\
Threshold: Homeowners ($L^{d,h=1}_\mathrm{I}$) & \$178,000 & \$252,500\\
Threshold: Non-homeowners ($L^{d,h=0}_\mathrm{I}$) & \$307,000 & \$381,500\\
Rate of Reduction ($\varpi^d_{\mathrm{A}}$) & \$0.039 & \$0.039\\
\hline
\end{tabular}
\end{table}

\subsection{Calibration model}
\label{sec:calibrationmodel}
Calibration of the model on the sample data is performed with the maximum likelihood method. The sample data is split into  vectors of single ($d=\mathrm{S}$) and couple ($d=\mathrm{C}$) households for consumption ($\mathbf{c}^d$) and housing ($\mathbf{h}^d$). Denote total data as $\mathcal{D} = \{\mathbf{c}^\mathrm{S}, \mathbf{c}^\mathrm{C}, \mathbf{h}^\mathrm{S}, \mathbf{h}^\mathrm{C}\}$. The statistical models are assumed to be $c^d_i = \widetilde{c}^d_i(\Theta) e^{\epsilon^d_{i}}$ and $h^d_i = \widetilde{h}^d_i(\Theta) e^{\varepsilon^d_{i}}$, where $c^d_i$ and $h^d_i$ are sample $i$ from the corresponding dataset, $\widetilde{c}^d_i(\Theta)$ and $\widetilde{h}^d_i(\Theta)$ are the optimal consumption and housing respectively from the model based on wealth and age corresponding to sample $i$ and the utility model parameters vector is
\begin{equation}
\Theta^\top = \left( \begin{array}{c c c c c c c c c c}
\gamma_\mathrm{S} & 
\gamma_\mathrm{C} &
\gamma_\mathrm{H} &
\theta &
a &
\overline{c}_\mathrm{S} &
\overline{c}_\mathrm{C} &
\psi &
\lambda \end{array} \right).
\end{equation}
Finally, $\epsilon^d_{i} \sim \mathcal{N} (0, \sigma^d_{\epsilon})$ and $\varepsilon^d_{i} \sim \mathcal{N} (0, \sigma^d_{\varepsilon})$ are the independent non-standardized error terms (residuals).

The log likelihood function of $N$ independent identically distributed samples $\mathbf{x} = (x_1, x_2, ..., x_N)$ from a log-normal distribution such that $\ln x_i \sim \mathcal{N}(\ln \mu, \sigma)$ is

\begin{equation}
\mathcal{L}_{\mathbf{x}}(\mu,\sigma) \propto  -\frac{N}{2} \ln \left( 2 \pi \sigma^2 \right) - \frac{1}{2 \sigma^2}  \sum _{i=1}^N  \left(\ln x_i - \ln \mu \right)^2.
\end{equation}
The maximum likelihood estimates of parameters $\Theta$ and $\bm{\sigma} = \{\sigma^\mathrm{S}_{\epsilon}, \sigma^\mathrm{C}_{\epsilon}, \sigma^\mathrm{S}_{\varepsilon}, \sigma^\mathrm{C}_{\varepsilon}\}$ are obtained by maximizing the total log likelihood
\begin{equation}
\mathcal{L}_{\mathbf{c}^\mathrm{S}}(\mathbf{\widetilde{c}}^\mathrm{S}(\Theta),\sigma^\mathrm{S}_\epsilon) + \mathcal{L}_{\mathbf{c}^\mathrm{C}}(\mathbf{\widetilde{c}}^\mathrm{C}(\Theta),\sigma^\mathrm{C}_\epsilon) + \mathcal{L}_{\mathbf{h}^\mathrm{S}}(\mathbf{\widetilde{h}}^\mathrm{S}(\Theta),\sigma^\mathrm{S}_\varepsilon) +  \mathcal{L}_{\mathbf{h}^\mathrm{C}}(\mathbf{\widetilde{h}}^\mathrm{C}(\Theta),\sigma^\mathrm{C}_\varepsilon),
\end{equation}
with respect to $(\Theta, \bm{\sigma})$.

Other distributions such as skew-t lead to some slight improvement in residual fitting, but since it is not very significant it has not been included in this paper.

\subsection{Calibrated parameters}
Our estimated parameters (see Table \ref{table:parameters}) are in line with related literature. The risk aversion is slightly lower than \cite{Ding2013} who estimated $\gamma=-3$, while the consumption floor is well under the full Age Pension rates but in line with the authors findings. This can be compared with \cite{Ameriks2011} where the consumption floor is only \$5,750 USD (the average social security payment), hence in relative terms our estimates are higher. 

\begin{table}[h]
\centering
\caption{Calibrated parameters with standard errors}
\label{table:parameters}
\begin{tabular}{l c c c c c c c c c}
\hline
& $\gamma_\mathrm{S}$ & $\gamma_\mathrm{C}$ & $\gamma_\mathrm{H}$ & $\theta$ & $a$ & $\overline{c}_\mathrm{S}$ & $\overline{c}_\mathrm{C}$ & $\psi$ & $\lambda$ \\
\hline
Value & -1.98 & - 1.78 & -1.87 & 0.96 & 20 726 & 10 122 & 15 702 & 1.18 & 0.044 \\
Std. Error & 0.38 & 0.37 & 0.37 & 0.01 & 208 & 1 648 & 1 826 & 0.03 & 0.009 \\
\hline
\end{tabular}
\end{table}

The calibrated health proxy parameter $\psi = 1.18$ indicates that the preferences for consumption which exceed the consumption floor decreases with a factor of $\psi^{(\gamma_d-1)/2}$, hence 0.61 for singles and 0.63 for couples each year due to declining health. To put this into perspective it equals a decrease of \$1,320 per year for the median single household between age 65-75, and \$2,312 for the median couple household. \cite{Bernicke2005} conducted a US study on empirical retirement data and found that the difference in consumption for the same age span was 26.4\%, and \cite{Higgins2011} find a 20-30\% drop in median levels where the decrease is larger for wealthier households, which confirms our results. \cite{Clare2014} finds an average decrease of 10\% for households with a comfortable lifestyle (roughly 300\% higher than our consumption floor) and 2\% for a modest lifestyle (roughly 100\% higher than our consumption floor) between the ages 70 to 90. Our calibrated model suggests similar decreases in expenditure, and captures the characteristics of larger declines for wealthier households.

Finally, $\theta$ implies little sensitivity of consumption to wealth, consistent with an explanation that additional wealth is being saved for a bequest. This is very similar to \cite{Ding2013} ($\theta = 0.956$), which indicates that a sequential model does not affect bequest motives. That said, our model cannot separate whether this is due to precautionary savings or indeed clear bequest motives.

\section{Results}

The calibration output indicates that the model fits the empirical behavior, and the statistical model is well chosen as the residuals have an acceptable Quantile-Quantile fit. The assumption that consumption and housing residuals are independent is confirmed as well. That said, due to the limitations in the data used for calibration (see Section \ref{sec:dataset}) the result should only be considered for healthy households in the post retirement phase. This is because the survey sample tends to be of better health than what might be true in reality, as retirees with bad health are more likely to live in an assisted care facility.

\subsection{Optimal Consumption}
\begin{figure}[!h]
\centering
\centerline{\includegraphics{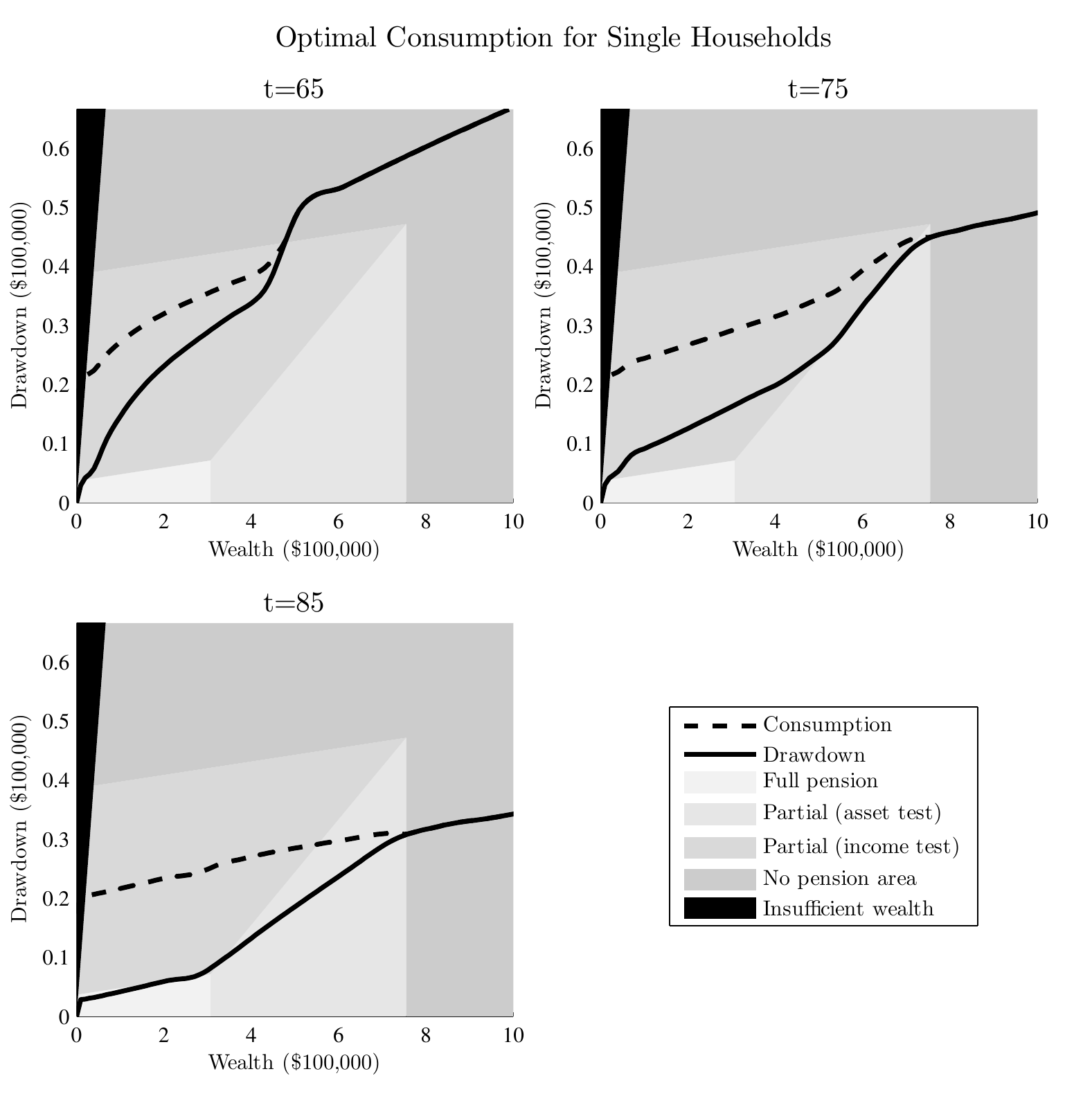}}
\caption{Optimal drawdown ($\alpha_t w_t$) and consumption in relation to wealth for the single non-homeowner case. The shaded zones show whether the asset or income test is binding and leading to partial Age Pension, or if the retiree would receive full/no Age Pension. The graphs illustrate the different drawdown and consumption curves for ages $t=$ 65, 75 and 85 years.}
\label{fig:graph_optdd_single}
\end{figure}

The optimal consumption curve in relation to wealth differs from the one in traditional utility models, where the deviations can be explained by the Age Pension means-test parameters (Figure \ref{fig:graph_optdd_single}). Traditionally, consumption is a smooth, concave and monotone function of wealth for risk averse agents. Generally this is true for drawdown outside the upper thresholds of the means-test where no Age Pension is received, but as the means-test binds the optimal drawdown policy changes slightly to anticipate the Age Pension received (which equals the area between the dashed and solid curve). Note that the drawdown curves in relation to the Age Pension thresholds are very similar between single and couple households. The rate of drawdown in relation to wealth decreases faster at the point where the retiree goes from full Age Pension to partial Age Pension due to the income test binding. This is effectively sacrificing current utility from consumption in order to receive additional future utility from consumption or bequest by adding Age Pension to liquid assets. This effect decreases with age however, as the sum of expected future Age Pension decreases due to mortality risk increasing, decreased consumption over one period will have a larger relative marginal utility loss.

\begin{figure}[!h]
\centering
\centerline{\includegraphics{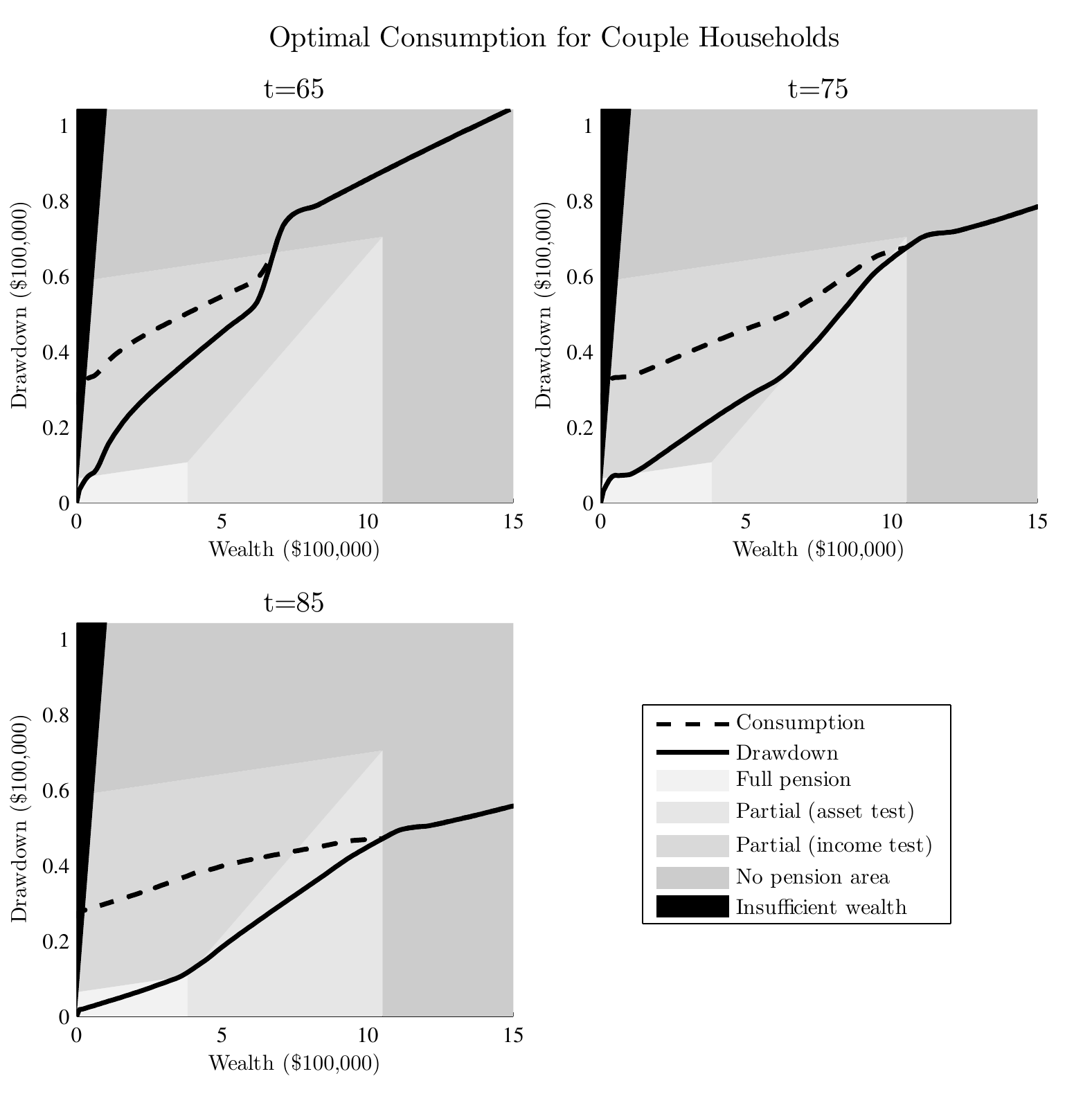}}
\caption{Optimal drawdown ($\alpha_t w_t$) and consumption in relation to wealth for the couple non-homeowner case. The shaded zones show whether the asset or income test is binding and leading to partial Age Pension, or if the retiree would receive full/no Age Pension. The graphs illustrate the different drawdown and consumption curves for ages $t=$ 65, 75 and 85 years.}
\label{fig:graph_optdd_couple}
\end{figure}

The most obvious effect of the Age Pension means-test on optimal drawdown can be seen between ages 75-85. The optimal drawdown curve almost perfectly follows the lower threshold of the income test, and continues where the income and asset tests intersect. Such behavior would maximize the Age Pension received for less wealthy households, until the curve tapers off into the asset test zone to avoid increased consumption. 

As the retiree ages, the consumption in relation to wealth flattens out even further to match the effect of declining health. At this point the optimal consumption shows almost no regard to the means-test, and converges towards the expected result from a utility model without means-tested Age Pension. For very low levels of wealth the drawdown curve disappears to negative territory, indicating that part of the Age Pension shall be saved as desired consumption can be fully covered by the Age Pension received.

\begin{figure}[!h]
\centering
\centerline{\includegraphics{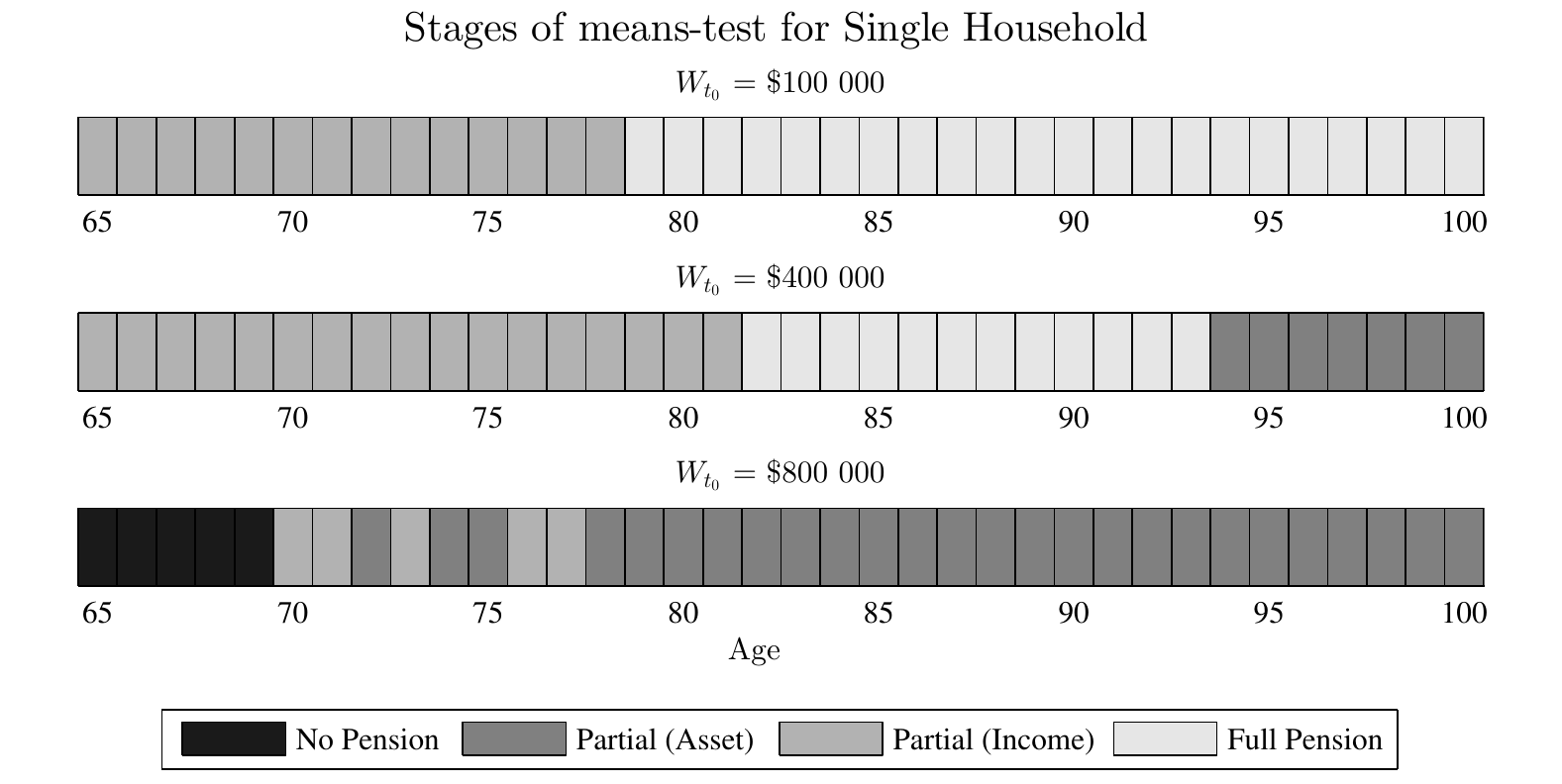}}
\caption{Phases in means-testing for expected wealth evolution and optimal drawdown paths given starting wealth $W_{t_0}$. Each rectangle corresponds to whether the means-test binds or if full Age Pension is received for the year.}
\label{fig:graph_stages}
\end{figure}

An interesting result of this is the sequence of the means-test phases. \cite{Ding2013} suggests that retirees go through the Age Pension in phases in the order of no pension, partial pension due to assets test, partial pension due to income test, and full pension. This is not necessarily true, especially when behavior such as higher consumption early in retirement and wealth accumulation for wealthier households are considered. 

It can be seen in Figure \ref{fig:graph_stages} that phases tend to vary for different expected wealth and drawdown paths from $t_0$ to $T$, where wealth grows with the expected return between each periods. Less wealthy households ($W_{t_0} = \$100,000$) tend to stay on full pension as the wealth is too low to grow to the asset test threshold, despite lower consumption with age. Wealthier households ($W_{t_0} = \$400,000$) however tend to accumulate enough wealth so that the asset test binds after a period of full Age Pension. Early in retirement it is possible to switch between phases, where risky returns can increase wealth so that the asset test binds, which in turn leads to a higher drawdown amount and binding of the income test.

\subsection{Optimal risky asset allocation}
\begin{figure}[!h]
\centering
\centering
\includegraphics{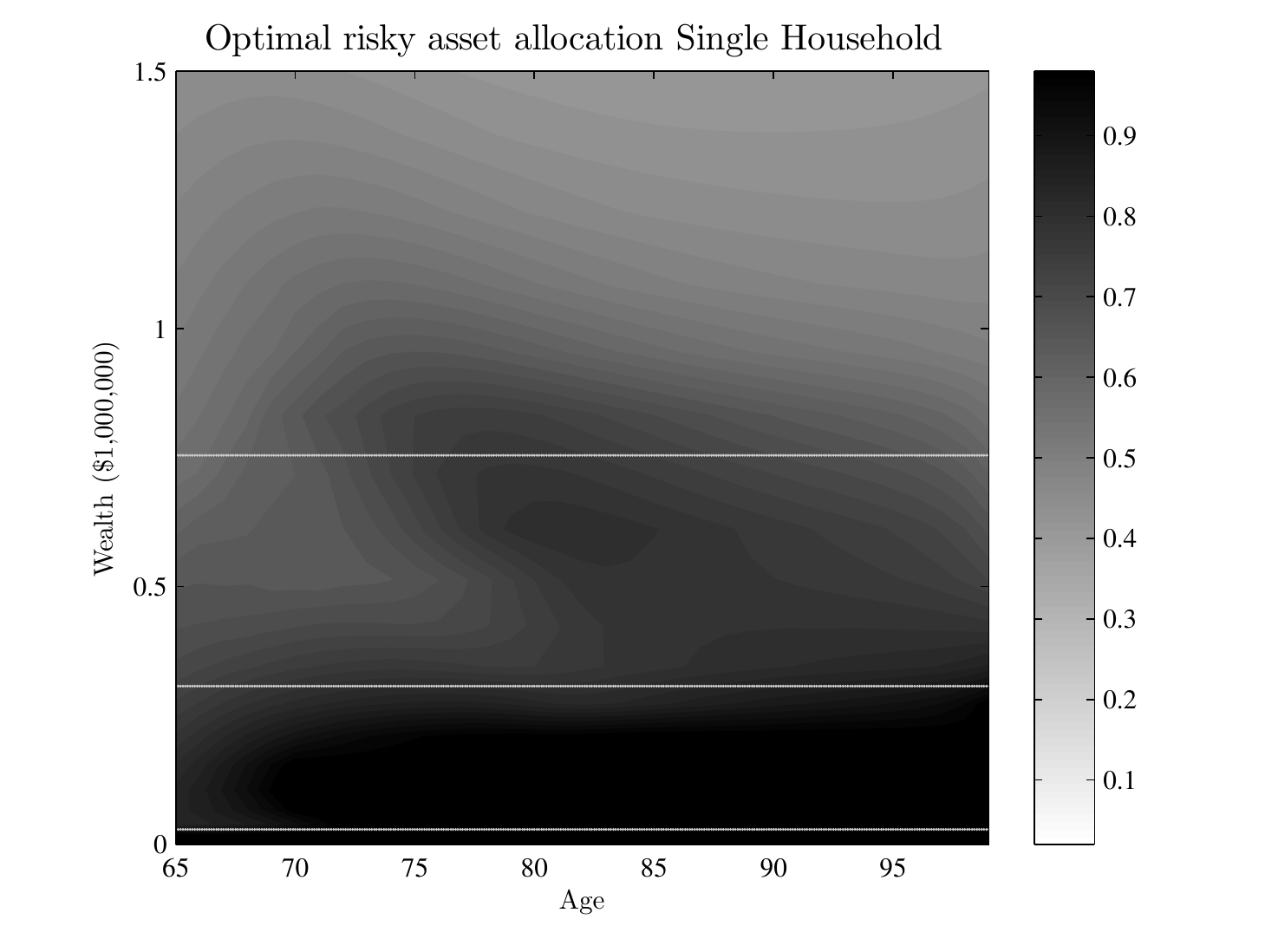}
\caption{Optimal allocation in risky assets for single non-homeowners. The horizontal lines (from bottom up) show the threshold $a$, the threshold for partial Age Pension due to asset test, and the threshold for no Age Pension due to asset test.}
\label{fig:graph_allocation}
\end{figure}

The exposure to risky assets in the portfolio is highly dependent on wealth and age. In general, the percentage allocation tends to increase with age for less wealthy households but decrease for wealthier ones. The increase in less wealthy households is contrary to traditional investment advice, which suggests that the allocation of risky assets should be reduced with age. The effect can be seen in Figure \ref{fig:graph_allocation}. This confirms the findings of \cite{Iskhakov2015} (as long as the consumption floor is less than the Age Pension) and \cite{Ding2013} who show that when bequest is considered a luxury, the optimal allocation of risky assets increases with age, implying higher allocation to risky asset throughout retirement. Only when bequest is not considered in a utility maximization model is decreasing the exposure with age indeed optimal \citep{Blake2014}. 

\begin{figure}[!h]
\centering
\centering
\includegraphics{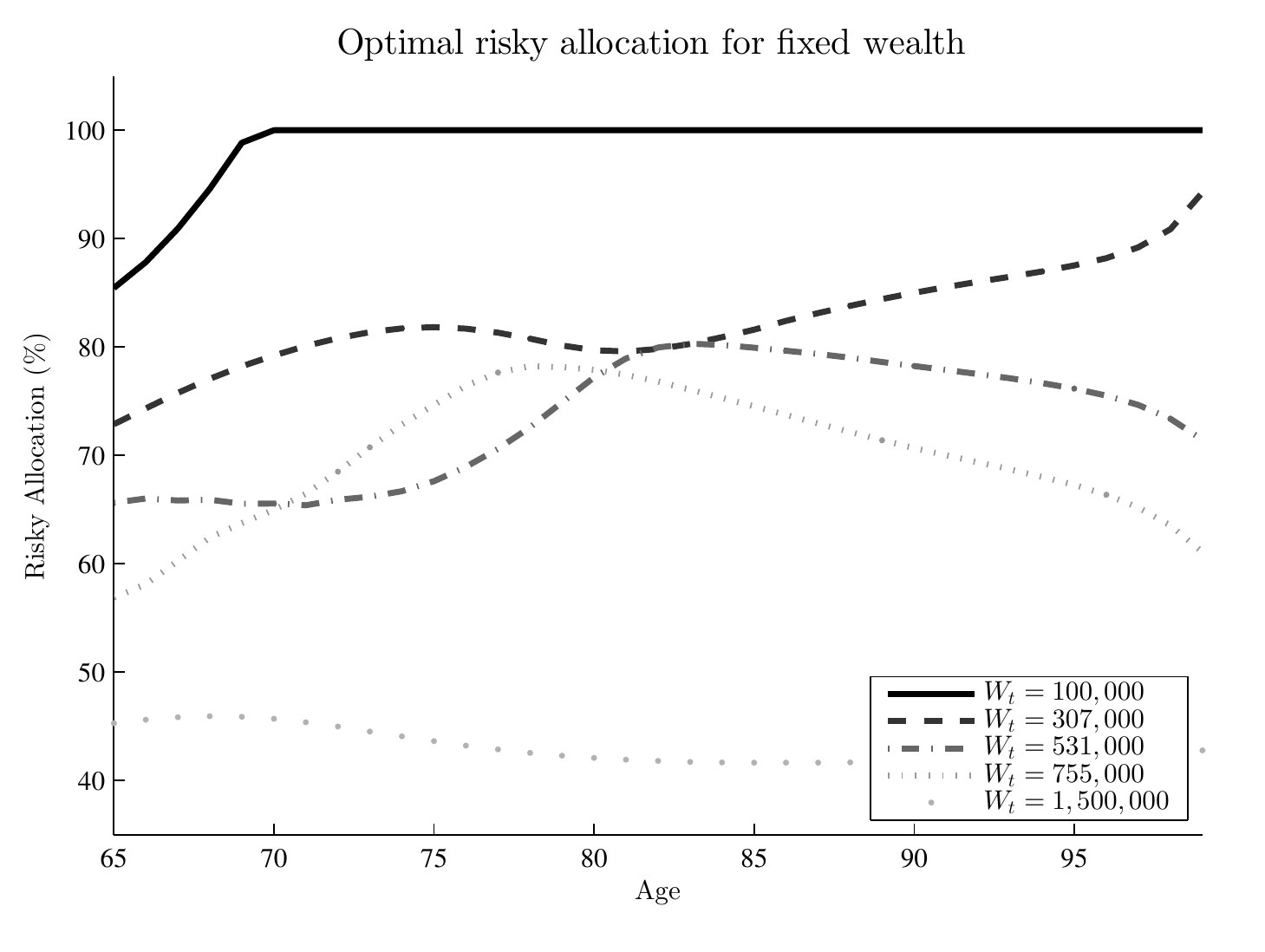}
\caption{Optimal allocation in risky assets for single non-homeowners, given a fixed wealth. The wealth levels correspond to full pension, the lower threshold of the asset test, partial pension due to asset test, upper threshold of the asset test and no pension.}
\label{fig:graph_risky_S_dissected}
\end{figure}

The contour chart shows a complex relationship with the means-test and Age Pension. Exposure tend to be more aggressive when the asset test is binding, as potential losses will be partly offset by increased Age Pension payments, hence Age Pension and risky assets are negatively correlated. The local maxima (dark area) in the middle corresponds with where this offset effect is the greatest (this will disappear if minimum withdrawals are enforced as studied in Section \ref{sec:MinWithdrawals}). The relative decrease vertically towards full pension is due to the buffer is no longer proportional, as larger losses cannot be compensated with more than full Age Pension. As the mortality risk increases, the expected buffer over the remaining life decreases hence allocation decreases horizontally towards old age. The downward slope in the local maxima is because of the optimal drawdown rules. Since drawdown decreases with age the risk of significantly decreasing wealth decreases as well, resulting in lower levels of buffer against losses being accepted. 
When no Age Pension is received from the income test, there is a higher allocation to risky assets the closer to partial pension we get. As with the asset test, this is due to losses being offset by Age Pension. The maximum increase is therefore on the threshold between no and partial Age Pension, as this is where the buffer has the greatest effect. \cite{Hulley2013} finds that risky allocation is much higher when the asset test is binding due to the steeper taper rate, and slightly lower (but still higher than the benchmark) for the income test. The effect of the Age Pension acting as a safety net to investment is further implied as lower wealth suggests full exposure to risky assets. These characteristics become clear in Figure \ref{fig:graph_risky_S_dissected}.

In addition to the effects from the Age Pension, a few general conclusions can be derived:
\begin{itemize}
\item[-] As wealth increases, allocation decreases. A loss of wealth has more negative marginal utility than the equivalent gain has positive utility. The difference increases with larger wealth.
\item[-] As the retiree ages, the (mortality risk) weight increases towards bequest where the marginal utility decreases quickly with increased wealth, but with a smaller negative derivative. This means that preserving capital becomes more important with age as wealth increases.
\item[-] Couples tend to be more aggressive with risky assets. One factor is the slightly lower risk aversion compared with singles, but since mortality risk is also lower the couple has a higher chance of recovering negative asset shocks, meaning they have more to gain from higher risk exposure.
\end{itemize}

\subsection{Optimal housing allocation}

\begin{figure}[!h]
\centering
\centerline{\includegraphics{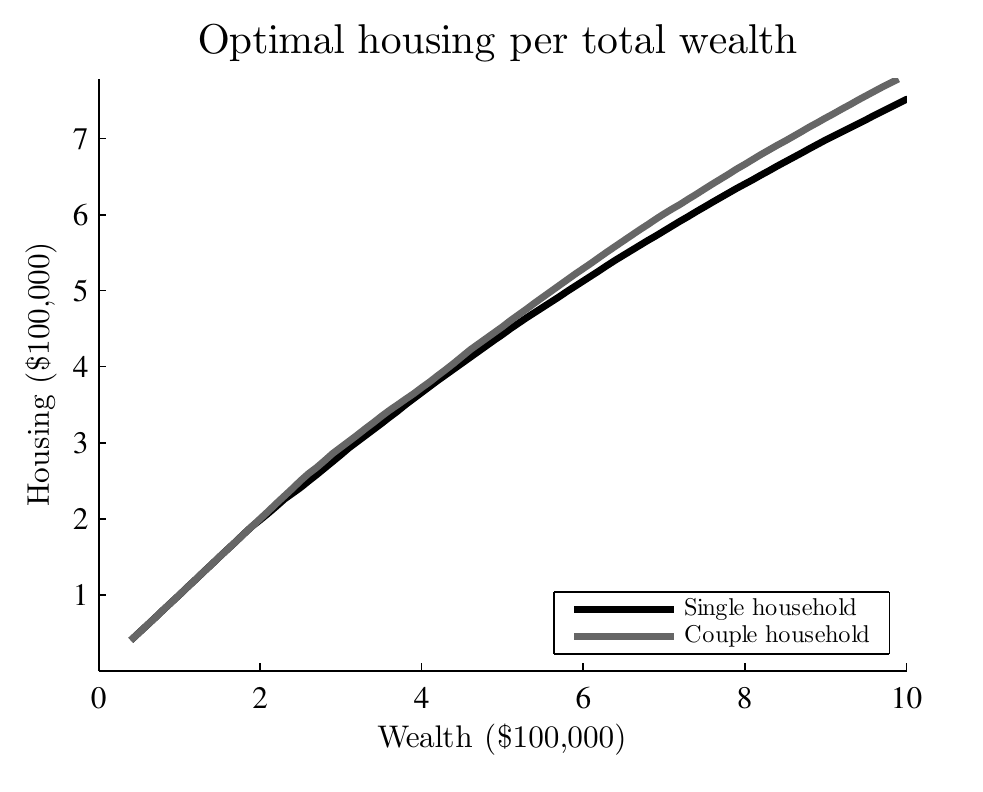}}
\caption{Optimal allocation of total wealth into housing for single and couple households.}
\label{fig:graph_housing}
\end{figure}

The literature is inconclusive regarding whether retirees allocate more assets into housing in order to be eligible for either full or partial Age Pension. We find no evidence that the model considers means-test levels for optimal allocation of net assets into housing, such as whether the asset test binds or not at $t_0$. This would show up as kinks in Figure \ref{fig:graph_housing} where the wealth equals the asset test thresholds.

The risk aversion for housing falls between the risk aversion of singles and couples. The different curves for optimal housing is then due to marginal utility of housing in relation to the marginal utility of consumption. As wealth increases, the risk aversion for couples favors housing more in relation to consumption than single households do. In addition, liquid assets for couples tend to be higher hence results in a higher dollar value allocated into housing than for single households.

\subsection{Forced minimum withdrawals}
\label{sec:MinWithdrawals}
Allocated pension accounts impose minimal withdrawal rates based on age (Table \ref{table:minwithdrawal}). It can be argued that this should be enforced in the calibration model, however we have intentionally left it out for the following three reasons:
\begin{itemize}
\item[-] Forced withdrawal does not necessarily affect the level of consumption, but it does affect whether the consumption consists of drawdown of the wealth or Age Pension received. A retiree that is forced to withdraw a larger amount of wealth than is optimal will effectively replace part of the Age Pension with own funds, but consume a similar dollar amount.
\item[-] In response to the global financial crisis of 2007-08, the government provided pension drawdown relief by reducing the withdrawal rates by 50\% between 2008-11 and 25\% between 2011-13. As the data set is taken from 2009-2010, these lower withdrawal rates applied.
\item[-] The benefit (income deduction) of an allocated pension account is no longer available for new accounts after 1\textsuperscript{st} of January 2015, hence new retirees might opt for a different kind of account. These accounts might not have forced minimal withdrawals.
\end{itemize}

\begin{table}[!h]
\centering
\caption{Minimum withdrawal rates for allocated pension accounts for the year 2013 and onwards.}
\label{table:minwithdrawal}
\begin{tabular}{l c c c c c c c}
\hline
Age & $\le$ 64 & 65-74 & 75-79 & 80-84 & 85-89 & 90-94 & 95 $\le$ \\
Min. drawdown & 4\% & 5\% & 6\% & 7\% & 9\% & 11\% & 14\% \\
\hline
\end{tabular}
\end{table}

\begin{figure}[!h]
\centering
\centerline{\includegraphics{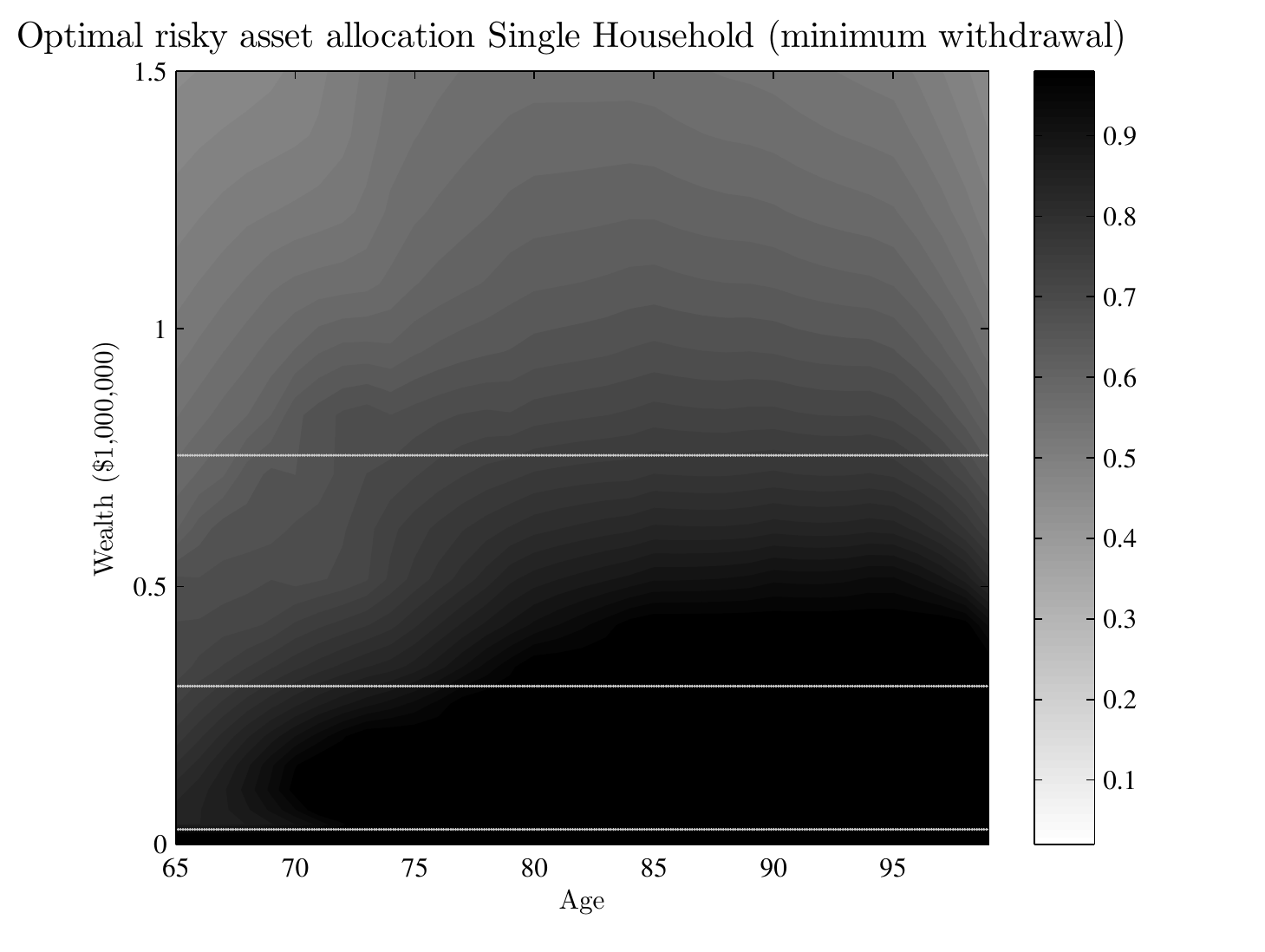}}
\caption{Optimal allocation in risky assets for single non-homeowners with forced minimum withdrawal. The horizontal lines (from bottom up) show the threshold $a$, the threshold for partial Age Pension due to asset test, and the threshold for no Age Pension due to asset test.}
\label{fig:graph_risky_mindd}
\end{figure}

The minimum drawdown rules do however have an effect on the optimal results that needs to be brought to attention. First, minimum drawdown limits the assets from increasing with age due to decreased consumption. The retiree can still get switching cases between partial pension due to asset test and income test early in retirement, but this voids the case where the asset test can bind at older age as in Figure \ref{fig:graph_stages}. It could potentially happen in case of very high risky returns, but it is not a scenario that is expected. Secondly, it limits the freedom for planning drawdown and risky asset allocation to optimize the Age Pension received. It is still possible to some extent, but it is less available as the wealth level increases. This effect can easily be seen with optimal risky asset allocation (Figure \ref{fig:graph_risky_mindd}) compared with when no rules withdrawal are enforced (Figure \ref{fig:graph_allocation}). The allocation tends to follow the general rules where allocation decreases with wealth, increases with age for less wealthy households and but decreases for wealthier households. The exception is the increase for wealthy households until age 80-85, which is due to mortality risk shifting the weight towards bequest as the retiree ages where this age corresponds to the tipping point.

\section{Conclusions}
In this paper we develop a sequential expected utility model for the decumulation phase of Australian retirees. The model can be considered a more realistic extension to \cite{Ding2013} where we introduce stochastic wealth, stochastic family status and a health status proxy, and relax prior assumptions of the order of means-tests phases. The model is defined as a stochastic control problem where we solve for optimal consumption, optimal risky asset allocation, and optimal housing. We solve the problem numerically using dynamic programming, and calibrate the utility parameters against data from \cite{Statistics2011} using the maximum likelihood method.

We find that the model explains the general behavior of Australian retirees, such as the declining consumption due to age (health), and can suggest optimal policy for consumption and risky asset allocation with respect to means-tested Age Pension. The optimal policy is found to be highly sensitive to the means-test early in retirement, hence the retiree can plan ahead to take advantage of Age Pension. The effect fades however with age, as the expected future Age Pension decreases due to increased mortality risk. The possibility to plan drawdown and risky asset allocation with respect to Age Pension is however limited when minimum withdrawal rules are enforced - especially for wealthier households.

Since the family status is sequential, it is of interest to understand the differences between single households and couple households. In general, we find that the two behave very similar with respect to wealth drawdown, allocation to risky assets and allocation to housing. The main difference is that couples can accept higher investment risk due to lower mortality risk, and tend to favor owning a home slightly more than single households.

The means-test is often assumed to go through different phases sequentially. We find that this is only true if minimum withdrawal rules are enforced. If not, it's likely that households accumulate wealth later in retirement due to decreased consumption, hence a retiree can go from receiving full Age Pension to partial due to the asset test.

The model lends itself to applications on both macro and micro scale. It can be used to forecast future Age Pension needs, implications of policy changes (or new financial products) in retirement behavior, and on a financial planning level once individual risk preferences have been estimated. It can easily be extended to suit the defined contribution pension system of other countries, or future changes in the Australia Superannuation.

\section*{Acknowledgment}
This research was supported by the CSIRO-Monash
Superannuation Research Cluster, a collaboration among CSIRO, Monash University, Griffith University, the University of Western Australia, the University of Warwick, and stakeholders of the retirement system in the interest of better outcomes for all. Alex Novikov acknowledges the support of the Australian Research Council’s Discovery Projects funding scheme (DP130103315). We thank Xiaolin Luo for valuable discussions and comments.

\appendix
\numberwithin{equation}{section}
\section{Separation of Housing utility}
\label{AppendixHousing}
Consider the value function in equation (\ref{eq:FinalValueFunction}), which shows the optimal expected value function conditional on information at time $t=t_0$ if we use policy $\pi$ up to $t=T-1$. The following shows that utility for housing can be separated into a decision problem for $H$ at time $t=t_0$ separate from the consumption and bequest utility, in order to avoid having house value as a state variable (thereby avoiding an additional dimension). The following is for a single household, but the approach is valid for couple households as well.

Introduce an alternative reward function without the housing utility
\begin{equation}
 \overline{R}_{t}(W_t,G_t,\alpha_t) = \left\{ \begin{array}{ll}
         U_C(C_{t},G_t,t), & \mbox{if $G_t = 1,2,$}\\
         U_B(W_t), & \mbox{if $G_t = 0,$}\\
         0, & \mbox{if $G_t = \Delta.$}\end{array} \right. 
\end{equation}
Since $\mathbb{E}^\pi_{t_0}$ is condition on the wealth process and family status, and utility is time-separably additive, we have
\begin{equation}
\begin{aligned}
\widetilde{V} & = \underset{H}{\max} \left[ \underset{\pi}{\sup} \: \mathbb{E}^\pi_{t_0} \left[\beta_{t_0,T} \widetilde{R}(W_T,G_T) + \sum_{t={t_0}}^{T-1} \beta_{t_0,t} R_{t}(W_t,G_t,\alpha_t,H) \mid W_{t_0}, G_{t_0}\right] \right] \\
& = \underset{H}{\max} \left[ \underset{\pi}{\sup} \: \mathbb{E}^\pi_{t_0} \left[ \vphantom{\sum_{t={t_0}}^{T-1}} \,_{t_0}p^{\mathrm{S}}_{T-1} \, \beta_{t_0,T} \widetilde{R}(W_T,G_T=1) \right. \right. \\
& \quad \left. \left. + \sum_{t={t_0}}^{T-1} \beta_{t_0,t} \left( \,_{t_0}p^{\mathrm{S}}_{t} \left( U_C(C_{t},G_t=1,t) + U_H(H, G_t=1) \right) + \,_{t_0}p^{\mathrm{S}}_{t-1} (1-p^{\mathrm{S}}_{t-1}) U_B(W_t) \right) \mid W_{t_0} \right] \right] \\
& = \underset{H}{\max} \left[ \underset{\pi}{\sup} \: \mathbb{E}^\pi_{t_0} \left[ \vphantom{\sum_{t={t_0}}^{T-1}} \,_{t_0}p^{\mathrm{S}}_{T-1} \, \beta_{t_0,T} \widetilde{R}(W_T,G_T=1) \right. \right. \\
& \quad \left. \left. + \sum_{t={t_0}}^{T-1} \beta_{t_0,t} \left( \,_{t_0}p^{\mathrm{S}}_{t} \left( U_C(C_{t},G_t=1,t)\right) + \,_{t_0}p^{\mathrm{S}}_{t-1} (1-p^{\mathrm{S}}_{t-1}) U_B(W_t) \right) \mid W_{t_0} \right] \right. \\
& \quad \left. + \mathbb{E}^\pi_{t_0} \left[ \sum_{t={t_0}}^{T-1} \beta_{t_0,t}  \,_{t_0}p^{\mathrm{S}}_{t} \, U_H(H,G_t=1) \mid W_{t_0} \right] \right] \\
& = \underset{H}{\max} \left[ \underset{\pi}{\sup} \: \mathbb{E}^\pi_{t_0} \left[\beta_{t_0,T} \widetilde{R}(W_T,G_T) + \sum_{t={t_0}}^{T-1} \beta_{t_0,t} \overline{R}_{t}(W_t,G_t,\alpha_t) \mid W_{t_0}, G_{t_0}\right] \right. \\
& \quad \left. + \sum_{t={t_0}}^{T-1} \beta_{t_0,t}  \,_{t_0}p^{\mathrm{S}}_{t} \, U_H(H,G_t=1) \right].
\end{aligned}
\end{equation}
Define a new value function $\overline{V}$ from reward function $\overline{R}$
\begin{equation}
\overline{V}(W_{t_0},G_{t_0}) = \underset{\pi}{\sup} \: \mathbb{E}^\pi_{t_0} \left[\beta_{t_0,T} \widetilde{R}(W_T,G_T) + \sum_{t={t_0}}^{T-1} \beta_{t_0,t} \overline{R}_{t}(W_t,G_t,\alpha_t) \right],
\end{equation}
and we have
\begin{equation}
\widetilde{V} = \underset{H}{\max} \left[ \overline{V}(\mathsf{W}-H,G_{t_0}) + \sum^{T-1}_{t=t_0} \beta_{t_0,t} \,\,_{t_0}p^{\mathrm{S}}_t \, U_H(H,G_t=1) \right].
\end{equation}
The decision for allocation to housing is made from total wealth, and the remaining liquid assets after the house value has been subtracted are then available for the stochastic control problem to optimize consumption and investments.

The separation of housing can also be proven with recursion on the value function in the Bellman equation in a similar manner.

\bibliographystyle{te}

{\footnotesize
\bibliography{bibliography}}

\end{document}